\documentclass[%
 reprint,
 amsmath,amssymb,
 aps,
pra,
]{revtex4-2}

\usepackage{graphicx}
\usepackage{dcolumn}
\usepackage{bm}
\usepackage{amsmath}
\usepackage{amssymb}
\usepackage{amsfonts}
\usepackage[dvipsnames]{xcolor}
\usepackage{soul}
\usepackage{physics}
\usepackage{graphicx}
\usepackage{float}
\usepackage{subfig}
\usepackage{mathtools}
\usepackage{hyperref}
\hypersetup{colorlinks=true,
linkcolor=blue, citecolor=blue}
\usepackage[version=4,arrows=pgf-filled,
textfontname=sffamily,
mathfontname=mathsf]{mhchem}
\newcommand{\be}{\begin{equation}}
\newcommand{\ee}{\end{equation}}
\newcommand{\bea}{\begin{eqnarray}}
\newcommand{\eea}{\end{eqnarray}}

\begin{document}

\preprint{APS/123-QED}

\title{Quantum droplets and
Schrödinger's cat states in atomic-molecular Bose-Einstein condensates}

\author{Leena Barshilia$^{1,2}$}
 
\author{Rajiuddin Sk$^2$}

\author{Prasanta K. Panigrahi$^2$}
 \email{pprasanta@iiserkol.ac.in}
\author{Avinash Khare$^3$}
 \affiliation{ $^1$Department of Physical Sciences, Indian Institute of Science Education and Research Kolkata, West Bengal 741246, India
}%
\affiliation{$^2$Center for Quantum Science and Technology, Siksha 'O' Anusandhan University, Bhubaneswar, Odisha 751030, India}%
\affiliation{%
 $^3$Department of Physics, Savitribai Phule Pune University, Pune 411007, India}%

\begin{abstract}
Explicit realization of quantum droplets, even and odd Schrödinger cat states is demonstrated in an atom-molecular Bose-Einstein condensate in the presence of interconversion and Kerr non-linear interactions. The crucial roles of both the $\chi^2$-type nonlinearity and chemical potential in the formation of these macroscopic quantum states are shown, where the atomic condensate is in the cat state, with the corresponding molecular wave packet being a quantum droplet. The physical mechanism for their creation and common origin is established to be the non-linearity-induced self-trapping potentials, governed by photoassociation or Feshbach resonance, with the Kerr-type nonlinearities playing subdominant roles. The coexisting and controllable atom and molecular droplets are shown to realize the atom-molecular squeezed state with profiles ranging from Gaussian to flat-top super-Gaussian form. The Wigner functions are exhibited revealing the cat states' phase space interference and squeezing of droplets.
\end{abstract}

\maketitle


\section{Introduction}
Macroscopic quantum states, such as the Schrödinger cat and their generalizations, have primarily been realized in optical systems. These states are not only pivotal in illustrating foundational quantum principles but also find significant applications in quantum sensing and metrology \cite{pirandola2018advances,wang2020quantum,zurek2001sub,giovannetti2006quantum,li2017cat,pezze2018quantum,toth2014quantum, ghosh2015mesoscopic,carr2009cold} and information processing \cite{vlastakis2013deterministically}. The non-classical squeezed state, closely related to the cat state, \cite{shanta1994unified} is another class of interesting states \cite{klauder2006fundamentals,agarwal2012quantum}. Beyond these, other types of localized wave packets, such as dark and bright solitons, have also been extensively studied \cite{hasegawa1989optical,lakshmanan2012nonlinear,kivshar2003optical}. Since the pioneering works of Hasegawa \cite{hasegawa1973transmission}, these solitons in both optical systems \cite{kivshar2003optical} and atomic ensembles \cite{pethick2008bose,kevrekidis2016solitons} have been observed in various systems, including atomic Bose-Einstein condensates (BECs). The mean-field dynamics of these systems are typically governed by the Gross-Pitaevskii(GP) equation, which, in a quasi one-dimensional geometry, is equivalent to the integrable nonlinear Schrödinger equation \cite{shabat1972exact}.

In both optical and atomic systems, the primary nonlinearity arises from the Kerr effect, which can be precisely controlled in atomic systems through Feshbach resonance techniques \cite{kevrekidis2003feshbach}. In BEC, optical analogues, such as grey solitons \cite{shomroni2009evidence} and soliton trains \cite{nguyen2017formation}, have been experimentally realized in repulsive and attractive nonlinear regimes, respectively. The bright soliton in the form of a hyperbolic secant function exists in the case of attractive nonlinearity and has been widely explored \cite{kevrekidis2016solitons}. Recent studies have demonstrated the crucial role of quantum fluctuations in stabilizing collapsing wave packets, leading to the formation of quantum droplets \cite{petrov2015quantum} under free-floating conditions \cite{katsimiga2023interactions,spada2024quantum}. These quantum corrections, originally described by Lee-Huang-Yang(LHY) \cite{lee1957eigenvalues}, introduce dimension-dependent nonlinear modifications to the mean-field dynamics. 
The coexistence of cubic and quadratic nonlinearities in such systems gives rise to a rich variety of phases, including supersolid states \cite{parit2021supersolid}, and distinct solitonic structures like kinks \cite{shukla2021kink}. These nonlinear systems
have also enabled the realization of atomic lasers and other macroscopic quantum states \cite{robins2013atom}. Recently, increasing attention has been directed towards atomic-molecular Bose-Einstein condensates (AMBECs) due to their enhanced stability and the feasibility of maintaining cooling processes for these composite systems \cite{zhang2021transition}  with molecular BEC being recently observed \cite{bigagli2024observation}. Molecular solitons \cite{drummond1998coherent} and recently dark and bright solitons and kinks and anti-kinks as well as periodic excitations have been realized in the mean-field description of AMBEC \cite{modak2024chemical,modak2022coherent}. Collisional behaviour of ultracold atomic gases have been investigated in \cite{weiner1999experiments,hutson2007molecular,petrov2001interatomic,julienne1989collisions}, Debnath et al. 
\cite{debnath2023dynamics}. It is to be noted that scattering behaviour is quite different from that of solitons in two component nonlinear Schrödinger equations \cite{radhakrishnan1997inelastic}.\\
In analogous to optical systems, we investigate the possibility of realizing the non-classical states in the form of Schrödinger cat and squeezed states in AMBEC. The cat state and its generalizations are well known for their characteristic phase-space interference patterns, responsible for their sub-Fourier sensitivity and sub-Planck structures \cite{zurek2001sub}.
Schrödinger cat states have become iconic in illustrating quantum nonlocality, a fundamental aspect of the superposition principle, and the measurement-induced collapse of the wave function. These non-classical states are represented by a superposition of two Gaussian functions, each shifted to indicate the localization of the wave function at two distinct spatial positions, similar to a particle in a double-well potential. Coherent states, such as those representing laser fields, are eigenstates of the annihilation operator $\hat{a}$, satisfying $\hat{a}|\alpha\rangle = \alpha|\alpha\rangle$, and are naturally associated with Gaussian wavefunctions characteristic of classical fields. In contrast, cat states are eigenstates of the operator $\hat{a}^2$, and can be expressed as:

$$|\text{Cat}_\pm\rangle = \mathcal{N}_\pm \left( e^{-\frac{(x - \alpha)^2}{2}} \pm e^{-\frac{(x + \alpha)^2}{2}} \right)$$

where the state is a superposition of two Gaussians centered at positions $\pm \alpha$. These states have been extensively studied in the optical domain, leading to their experimental realization in laboratories. In optics, cat states can emerge in Kerr media, where nonlinearity plays a key role \cite{tara1993production}. This nonlinearity has also facilitated the observation of cat states in superconducting qubits, where circuit-QED architectures amplify the effects of nonlinearity \cite{he2023fast}. As is well known, Kerr nonlinearity in BEC has led to the formation of macroscopic, stable wave packets such as bright and dark solitons, represented by $\text{sech}$ and $\tanh$ functions, respectively.

Recently, analogues of atomic lasers and squeezed states have been explored due to their relevance in both foundational quantum mechanics and technological applications. In this work, we demonstrate the exact construction of odd and even cat states composed of bright solitons in an atomic-molecular Bose-Einstein condensate (AMBEC), where the key mechanism is the atom-molecular conversion process. We derive exact analytical expressions that transparently elucidate the roles of chemical potential, interconversion dynamics, and atom-atom as well as atom-molecule scattering processes identifying the key controlling parameters. We show that quantum droplets, interpreted as bound states of kink-anti-kink solitons, can exist in these systems, analogous to the bound states of bright solitons for even and odd cat states. All three non-classical states of the non-linear mean field equation describes the AMBEC are linear superposition of bright and dark solitons propagating non-linear excitations on atomic media have earlier shown to be superposition or more fundamental waves, leads to extreme enhancement and reduction of intensity due to ineterference effets \cite{panigrahi2003existence}.

The paper is organized as follows: in Sec. II, we describe the mean-field dynamics and highlight the physical origins of different nonlinearities. In Sec. III, we present the realization of quantum droplets and formulation of even and odd cat states and their corresponding phase-space behaviour. In Sec. IV, we summarize the conclusions and discuss some of the open problems. In Appendix \ref{Appendix_A}, we have provided the details of the three solutions discussed in Sec. III with the solutions of the consistency conditions are given. The expressions of self-consistent potentials are given in Appendix \ref{Appendix_B} which physically explain the origin of quantum droplets and cat states.  
\section{Model}

In the following, we briefly highlight the features of the mean-field description of AMBECs, which is a generalized two component nonlinear Schrödinger equation with a quadratically non-linear $\chi^2$-type interconversion term. This term can be controlled through photoassociation processes or Feshbach resonance. The mean field description takes into account the atom-atom, atom-molecule and molecule-molecule scatterings through Kerr-type cubic non-linearities. In a one-dimensional setting, this combination of quadratic and cubic nonlinearities gives rise to solitons whose classical and quantum properties are well-investigated .
Physically, interconversion involves the formation of molecules from two atoms
through the absorption of a photon during atomic collisions.
In terms of chemical reactions, it represents a second-order reaction:
\begin{equation}
\hat{H} =E_{A}\hat{n}_A +E_{A_2}\hat{n}_{A_2} + k(\hat{a}_A^\dag\hat{a}_A^\dag\hat{a}_{A_2} + h.c.),
		\label{ref{1}}
\end{equation}
where ground-state energies are denoted by $E_{A}$ and $E_{A_2}$.

It is worth mentioning that recent progress in the experimental front has allowed precise control of the internal degrees of freedom \cite{croft2017universality,krems2008cold,balakrishnan2016perspective} within the molecules in the domain of ultracold chemistry \cite{bohn2017cold,bell2009ultracold}, where photoassociation is playing a key role \cite{jones2006ultracold}. \\
 The mean field Hamiltonian is given by 
 \begin{eqnarray}
	\hat{H} &=& {\int} {\rm d}^3 r \Big( \hat{\psi}^{\dagger}_a \Big[ -\frac{\hbar^2}{2}\nabla^2 
	+ V_a^{trap}(\vec r) + \frac{g_a}{2} \hat{\psi}^{\dagger}_a \hat{\psi}_a \Big] \hat{\psi}_a \nonumber \\
	&+& \hat{\psi}^{\dagger}_m \Big[ -\frac{\hbar^2}{4}\nabla^2  
	+ V_m^{trap}(\vec r) + \epsilon + \frac{g_m}{2} \hat{\psi}^{\dagger}_m \hat{\psi}_m \Big] \hat{\psi}_m \cr \nonumber \\
	&+& g_{am} \hat{\psi}^{\dagger}_a  \hat{\psi}_a  \hat{\psi}^{\dagger}_m  \hat{\psi}_m
	+ \frac{\alpha}{\sqrt{2}} \big[ \hat{\psi}^{\dagger}_m \hat{\psi}_a \hat{\psi}_a 
	+ \hat{\psi}_m \hat{\psi}^{\dagger}_a \hat{\psi}^{\dagger}_a \big] \Big)
	\label{ref{2}}	
\end{eqnarray}
Here the field operators represent the resonant states of atoms and molecules, where $g_a, g_m$ and $g_{am}$ respectively quantifying the interaction strengths. This Hamiltonian describes a reversible reaction due to its hermiticity. 

In the following, the mean field equations are analyzed in a hydrodynamic approach, considering a quasi one-dimensional geometry. Using the normalization of refs.  \cite{salasnich2002effective, kamchatnov2004dynamics}, the mean-field equations are given by :

\begin{eqnarray}
i\frac{\partial\psi_{a}}{\partial t} &=& -\frac{1}{2}\frac{\partial^{2} \psi_{a}}{\partial x^{2}} + (V_a^{trap}+g_{a}|\psi_{a}|^{2} + g_{am} |\psi_{m}|^{2}) \psi_{a}\nonumber \\ &+& \alpha \sqrt{2}\psi_{m}\psi_{a}^{*}, \label{3} \nonumber\\ \\
i\frac{\partial\psi_{m}}{\partial t} &=& -\frac{1}{4}\frac{\partial^{2} \psi_{m}}{\partial x^{2}}+(V_m^{trap}+\epsilon +g_{m}|\psi_{m}|^{2}+g_{am}|\psi_{a}|^{2})\psi_{m} \nonumber \\ &+& 
\frac{\alpha}{\sqrt{2}}\psi_{a}^{2} \label{4}. \nonumber\\
\end{eqnarray}

Here $\psi_a$ and $\psi_m$ represent atomic and molecular field variables and $\alpha$ is the strength of interconversion. For the general scenario, we have incorporated the trap potentials  $V_a^{trap}$ and $V_m^{trap}$. Parameter $\epsilon$ represents the energy difference between the atoms and molecules in the conversion process.\\
It is noted that the present Hamiltonian respects the total particle number conservation.

\begin{equation}
N=\int{(\vert\psi_{a}\vert^2+2\vert\psi_{m}\vert^2}) dx=N_{a}+2N_{m}.
\end{equation}

In our study, we considered atomic-molecular BEC without the harmonic traps which can be incorporated through a similarity and gauge transformation \cite{atre2006class}.

\section{Exact solutions of AMBEC as linear superpositions of non-linear excitations}

The mean-field equations of AMBEC admit exact solutions, which are linear superposition of kink-anti-kink and displaced bright solitons. Interestingly, the combination of kink-anti-kink leads to the quantum droplet solution which has attracted considerable attention in the recent literature of atomic BEC. As mentioned earlier the presence of LHY correction term plays a key role in its occurrence. It is worth noting that quantum droplets are shown to form without trapping potential. Hence, we consider the solutions of the coupled Eqs. (\ref{3}) and (\ref{4}) in the absence of the potentials $V_a^{trap}$ and $V_m^{trap}$.
Experimentally also one can realize absence of trap by box potentials.\\

{\bf {Solution I}: Droplets in atomic and molecular condensates }\\

The exact solutions to the coupled Eqs. (\ref{3}) and (\ref{4}) has the form ~~$\psi_a = \dfrac{A}{B+\cosh^2(\beta x)} e^{-i\mu t}= A\psi^{spt}_a(x)e^{-i\mu t}$ ~~and~~ $\psi_m = \dfrac{D}{B+\cosh^2(\beta x)} e^{-2i\mu t}=D\psi^{spt}_m(x)e^{-2i\mu t}$,  provided certain relations are satisfied, which are given in the Appendix \ref{Appendix_A}. {$\psi_a^{spt}$ and $\psi_m^{spt}$, both of these wavefunctions can be expressed as the linear superposition of kink and anti-kink \cite{khare2022superposed}:

\be\label{psi_m_spt}
\psi_a^{spt}=\psi_m^{spt}=\frac{\tanh(\beta x +\Delta)- \tanh(\beta x -\Delta)} {\sinh(2\Delta)}
\ee
where $B = \sinh^2(\Delta)$. These atomic and molecular wavefunctions can be cast in a form which explicitly show the range of chemical potential in which the solutions manifest \cite{petrov2016ultradilute}.
We express both atomic and molecular wave functions in the form as given by Petrov.
With $\beta^2 = -\dfrac{\mu}{2}$, we can reexpress the solutions as: 
\be\label{P_9}
\psi_a = \frac{\sqrt{n_a}\frac{\mu}{\mu_0}}{
[1+\sqrt{1-\frac{\mu}{\mu_0}}\cosh(\sqrt{-8\mu} x)]} e^{-i\mu t}\
\ee 
\be\label{P_10}
\psi_m = \frac{\sqrt{n_m}\frac{\mu}{\mu_0}} 
{[1+\sqrt{1-\frac{\mu}{\mu_0}}\cosh(\sqrt{-8\mu} x)]} e^{-2i\mu t}\
\ee \\

where
\be\label{P_11}
\sqrt{n_a} = \frac{A(2B+1)}{2B(B+1)}, ~~\sqrt{n_m} = \frac{D(2B+1)}{2B(B+1)}\,,~~\frac{\mu}{\mu_0} = 
\frac{4B(B+1)}{(2B+1)^2}\,.
\ee\\

Fig. (\ref{fig:1a}) shows the Gaussian-like solitonic behaviour to the flat-top nature of the axial density profile of a quantum droplet as the chemical potential is varied. The flat-top transition occurs when $\mu\rightarrow\mu_0$, where $\mu_0$ is the critical value of the chemical potential:
\be\label{mu_0}
\mu_{0} = -\dfrac{4}{9}\dfrac{\alpha^2}{(g_a + g_{am})}
\ee
which depends on $\alpha$ and the interaction strengths. The relation between the interaction strengths is given by $g_m=\dfrac{(g_a-g_{am})}{2}$.

The quantum droplet exists for any region, $0< |\mu|< |\mu_{0}|$. This is akin to the case of atomic droplets. Eq. (\ref{mu_0}) clearly depicts that the range of the chemical potential is directly controlled by photoassociation or Feshbach resonance. It is to be noted here that these density profiles of atomic and molecular wavefunctions are identical. The physical reason for the droplet structure, that can be varied from Gaussian to super-Gaussian form and can be understood from the self-consistent potential when one expressed the droplet equation in the form of an eigenvalue equation. 
 The self-consistent potential corresponding to different density profiles are given in Figs. (\ref{fig:1b}) and (\ref{fig:1c}) ranging from box type (for flat top) to harmonic potential (Gaussian profile). The physical fact that the droplet is composed of a linear superposition of kink and anti-kink is clearly manifested in phase space. To illustrate this, the Wigner function profile will be plotted.
 Fig. (\ref{fig:3}) shows the Wigner function profile for the atomic as well as molecular BEC. It clearly shows the linear superposition in this nonlinear system in the form of interference. Further, one can observe the asymmetry of quadratures in the phase space, indicating the squeezed nature of these states.  \\

 \begin{widetext}
\begin{minipage}{\linewidth}
    \begin{figure}[H]
        \centering
        \subfloat[]{\includegraphics[width=0.32\textwidth]{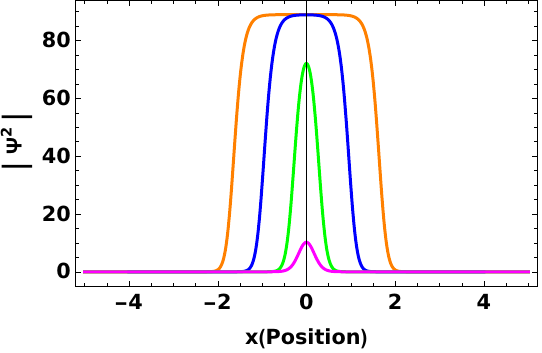}\label{fig:1a}}
        \subfloat[]{\includegraphics[width=0.32\textwidth]{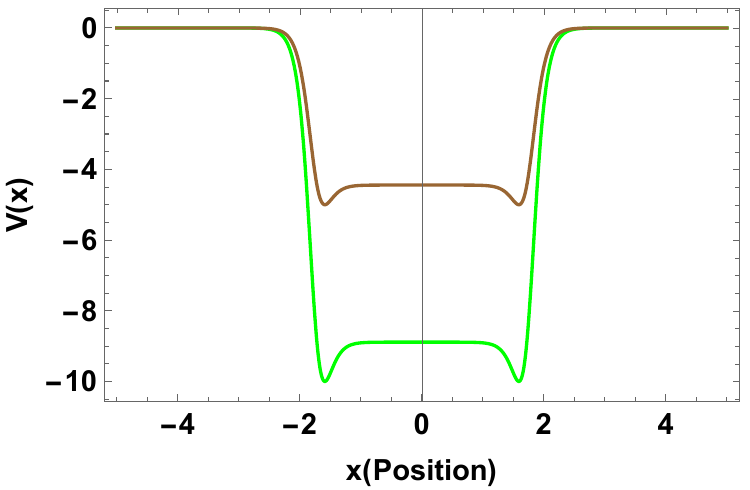}\label{fig:1b}}
        \subfloat[]{\includegraphics[width=0.32\textwidth]{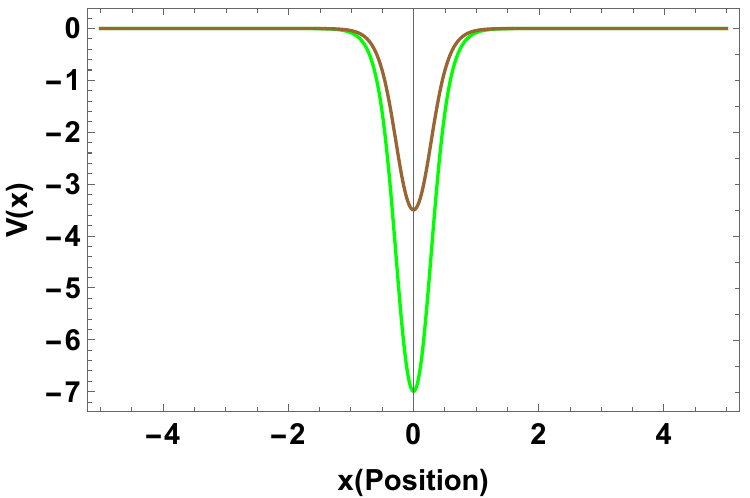}\label{fig:1c}}\\
         \subfloat[]{\includegraphics[width=0.25\textwidth]{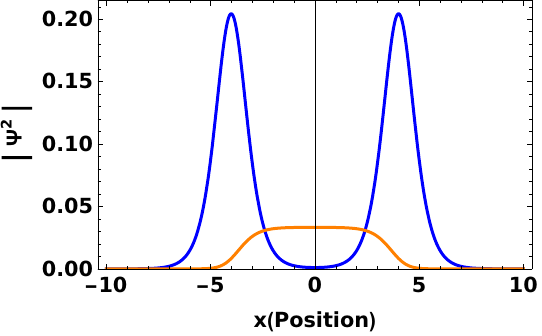}\label{fig:1d}}
         \subfloat[]{\includegraphics[width=0.245\textwidth]{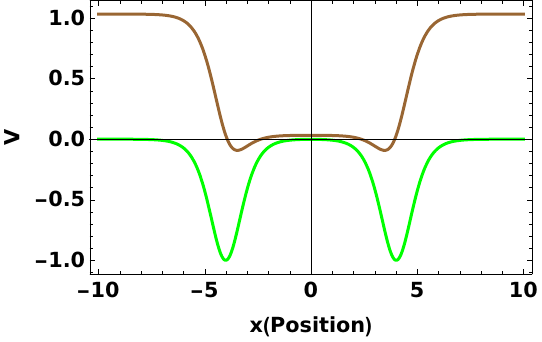}\label{fig:1e}}
         \subfloat[]{\includegraphics[width=0.25\textwidth]{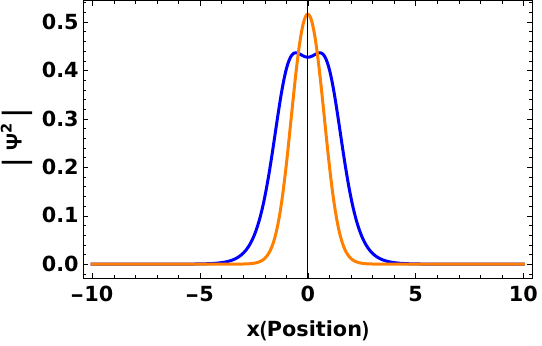}\label{fig:1f}}
         \subfloat[]{\includegraphics[width=0.245\textwidth]{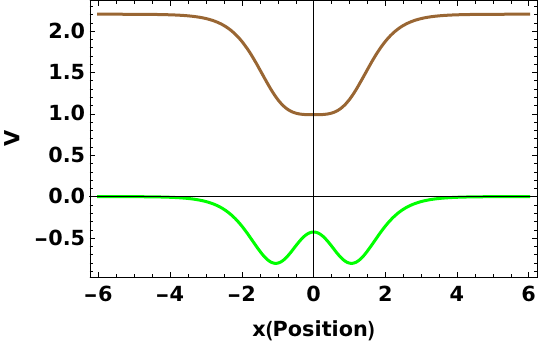}\label{fig:1g}}\\
         \subfloat[]{\includegraphics[width=0.25\textwidth]{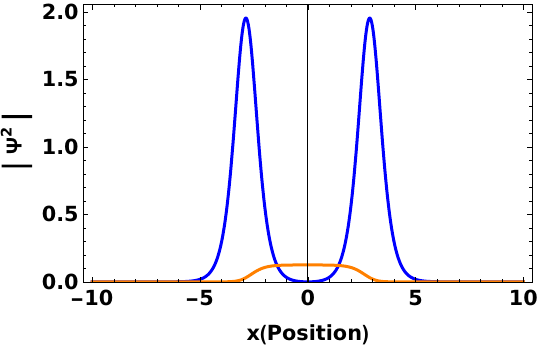}\label{fig:1h}}
         \subfloat[]{\includegraphics[width=0.245\textwidth]{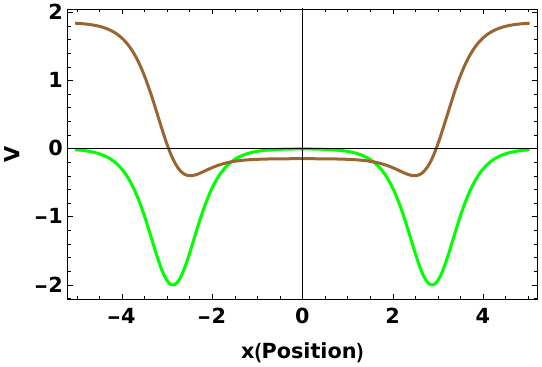}\label{fig:1i}}
          \subfloat[]{\includegraphics[width=0.25\textwidth]{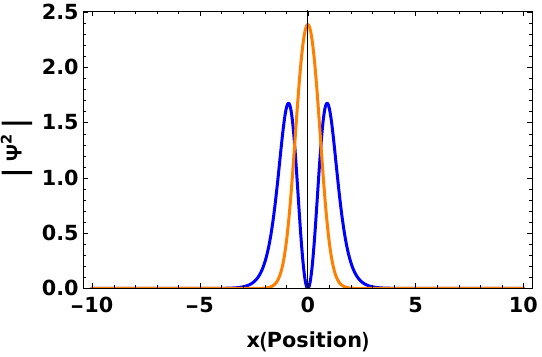}\label{fig:1j}}
           \subfloat[]{\includegraphics[width=0.245\textwidth]{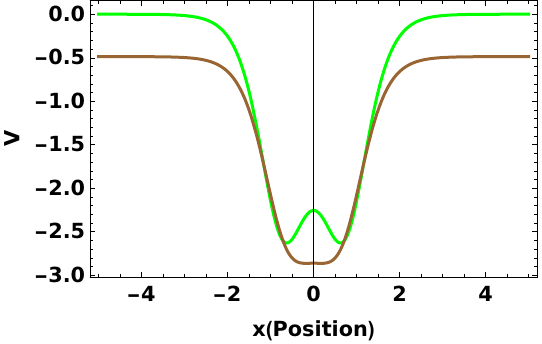}\label{fig:1k}}
        \caption{(a) Variation in density profiles for different $\mu$ for quantum droplet for parameters $g_a=3, g_m=2.9, g_{am}=-2.8$. $\alpha$ is chosen as 2, and $\mu_{0}=-\dfrac{80}{9}$ for the parameters used. As $\mu \rightarrow \mu_{0}$, the density at the center gradually increases and a flat plateau region is observed (orange curve).
        (b) Self-consistent potentials for atomic (green) and molecular (brown) condensates for the flat plateau region as depicted by orange color plot in Fig. (\ref{fig:1a}).
        (c) Self-consistent potentials for atomic (green) and molecular (brown) condensates for Gaussian profiles as depicted by pink color plot in Fig. (\ref{fig:1a}).\\
        Fig: (d)$\And$(f) show density profiles (blue for atomic BEC and orange for molecular BEC) for Solution II for the parameter sets ($g_a=-5$, $g_m=1$, $g_{am}=-2.41$)$\And$    ($g_a=-5$, $g_m=1$, $g_{am}=-1.1$)
        respectively and Fig: (e)$\And$(g) show their corresponding self-consistent potentials (green for atomic BEC and brown for molecular BEC).\\
        Fig: (h)$\And$(j) show density profiles (blue for atomic BEC and orange for molecular BEC) for Solution III for the parameter sets ($g_a=-1.03$, $g_m=-1.2$, $g_{am}=-0.53$)$\And$  ($g_a=-1.03$, $g_m=-1.2$, $g_{am}=-.8$)
        respectively and Fig: (i)$\And$(k) show their corresponding self-consistent potentials (green for atomic BEC and brown for molecular BEC).\\
        (Expressions for self-consistent potentials are given by Eqs. (\ref{B5}) and (\ref{B6}) in Appendix \ref{Appendix_B})} 
        \label{Fig_1}
    \end{figure}
\end{minipage}
\end{widetext}

{\bf {Solution II} : Atomic even cat state and
molecular squeezed state}

The linear superposition of displaced bright solitons are in the form of cat states . In the following, we give explicit solutions to the cat state for the coupled system. The atomic solution is in the even cat state form, and the molecular solution is in the squeezed state. The exact solution to the coupled Eqs. (\ref{3}) and (\ref{4}) has the form ~~$\psi_a = \dfrac{A\cosh(\beta x)}{B+\cosh^2(\beta x)} e^{-i\mu t}=A\psi^{spt}_a(x)e^{-i\mu t}$ ~~and~~ $\psi_m = \dfrac{D}{B+\cosh^2(\beta x)} e^{-2i\mu t}=D\psi^{spt}_m(x)e^{-2i\mu t}$ ~~ provided certain relations are satisfied, which are given in Appendix \ref{Appendix_A}. Fig. (\ref{fig:1d}) and Fig. (\ref{fig:1f}) show the density plots for the atomic(ground state) and molecular condensates for the flat molecular profile to the Gaussian molecular profiles by varying the interaction strengths.   
Fig. (\ref{fig:1e}) shows the symmetric double well type potential with high middle barrier and Fig (\ref{fig:1g}) shows the symmetric double well type potential with low middle barrier, which leads to the cat state  localized in two spatially separated positions for atomic case whereas for molecular case it is showing the change from the box type potential to harmonic potential as also observed in droplets case.
It is to be noted that $\psi_a^{spt}$ can be expressed as the following superposition \cite{khare2022superposed}:
\be
\psi_a^{spt}=\frac{\sech(\beta x+\Delta)+\sech(\beta x -\Delta)}{2\cosh(\Delta)}
\label{EQ_10}
\ee

where $B = \sinh^2(\Delta)$.
Physically, this suggests that the atomic profile corresponds to a superposition of bright solitons. 
The $\psi_m^{spt}$ of the molecular BEC profile is the superposition of kink and anti-kink and has the same form as given in Eq. ({\ref{psi_m_spt}).
Fig. (\ref{fig:2a}) shows that the atomic BEC exhibits behaviour similar to an even cat state, as seen in optical systems. In contrast, the molecular profile behaves like a squeezed state, as shown in Fig. (\ref{fig:3}).
\begin{widetext}
\begin{minipage}{\linewidth}
\begin{figure}[H]
    \centering
\subfloat[]{\includegraphics[width=0.46\textwidth]{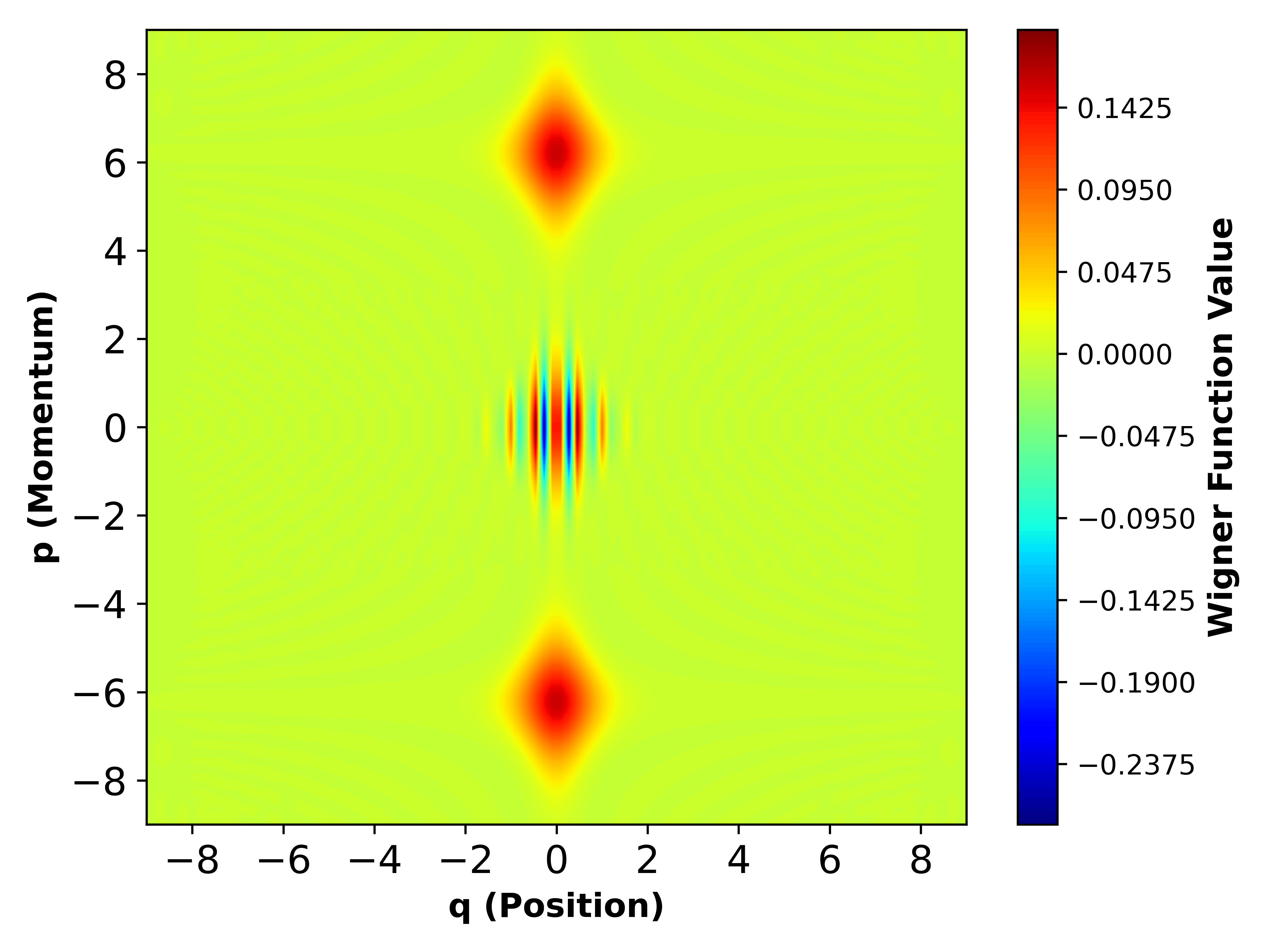}\label{fig:2a}}
\subfloat[]{\includegraphics[width=0.46\textwidth]{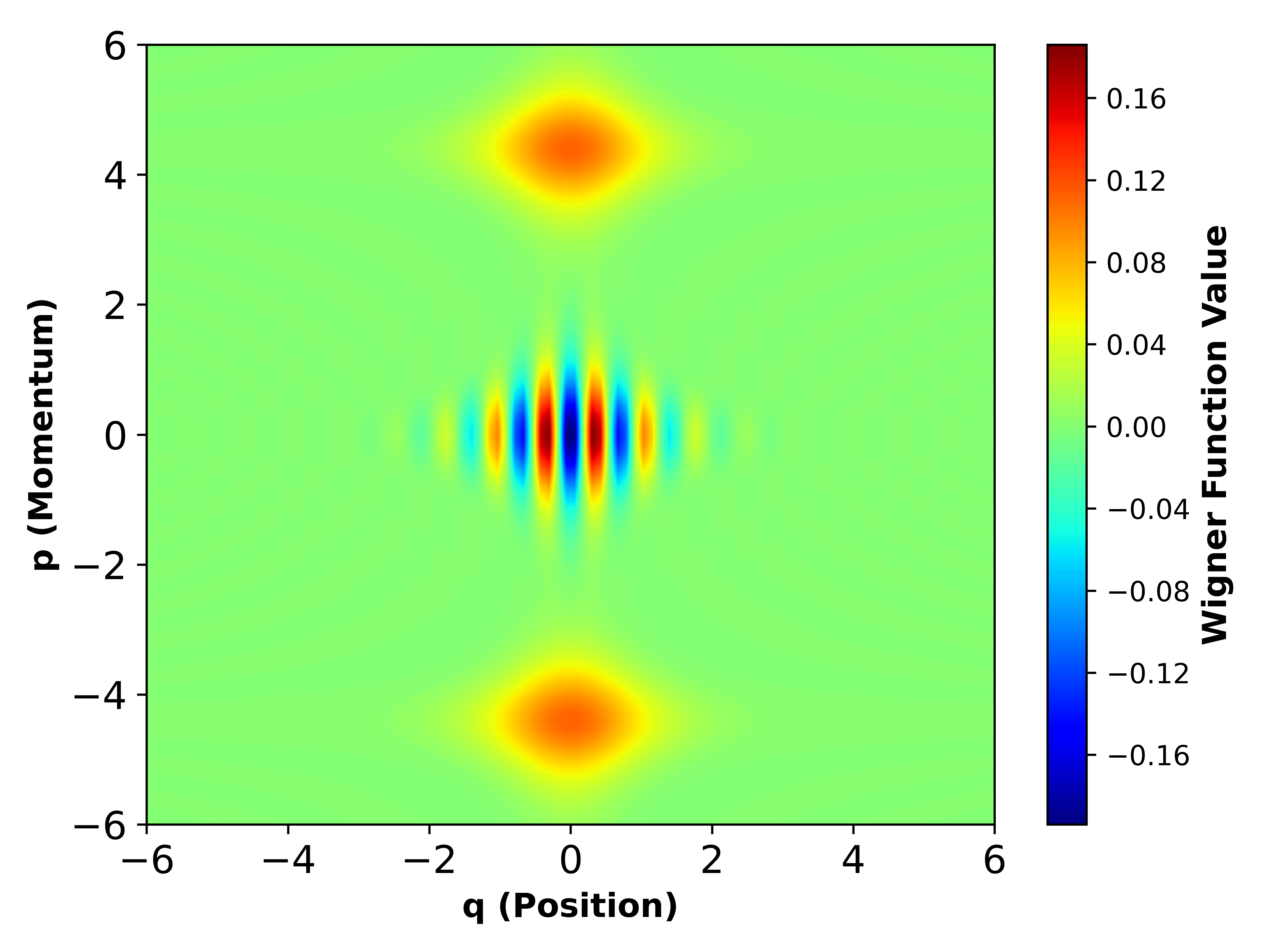}\label{fig:2b}}
\caption{Wigner function plot of the atomic (a) even cat state with the parameters $\beta=1$ and $\Delta=6.219$ and (b) odd cat state with the parameters $\beta=1.414$ and $\Delta=6.21077$. The atomic cat state is the superposition of bright solitons.}
\label{Fig_11}
\end{figure} 
\end{minipage}
\end{widetext}

{\bf {Solution III} : Atomic odd cat state and molecular squeezed state}\\

Another exact solution to the coupled 
Eqs. (\ref{3}) and (\ref{4}) has the 
form ~~$\psi_a = \dfrac{A\sinh(\beta x)}{B+\cosh^2(\beta x)} e^{-i\mu t}=A\psi^{spt}_a(x)e^{-i\mu t}$ ~~and~~ $\psi_m = \dfrac{D}{B+\cosh^2(\beta x)} e^{-2i\mu t}=D\psi^{spt}_m(x)e^{-2i\mu t}$~~  provided certain relations are satisfied, which are given in the Appendix \ref{Appendix_A}.  Fig. (\ref{fig:1h}) and Fig. (\ref{fig:1j}) show the density plots for the atomic(excited state) and molecular condensates for the flat molecular profile to the Gaussian molecular profiles by varying the interaction strengths. Fig. (\ref{fig:1i}) shows the symmetric double well type potential with high middle barrier and Fig (\ref{fig:1k}) shows the symmetric double well type potential with low middle barrier for atomic case whereas for molecular case it is showing the change from the box type potential to harmonic potential.

We note that $\psi_a^{spt}$  of the atomic BEC profile can be expressed in the following superposition form \cite{khare2022superposed}: 
\be
\psi_a^{spt}=\frac{\sech(\beta x-\Delta) - \sech(\beta x +\Delta)}{2\sinh(\Delta)}
\ee
 with $B = \sinh^2(\Delta)$. This represents a superposition of bright solitons with opposite parity, leading to the excited state. 
 The $\psi_m^{spt}$ of the molecular BEC profile is the superposition of kink and anti-kink and has the same form as given in Eq. (\ref{psi_m_spt}).
 
Akin to the quantum droplets, the chemical potential plays a significant role in the present case as well. Positivity of $\beta^2$ requires $\mu$ to be negatively bounded below by $\mu_0=0$, at which point the solution ceases to exist. 

 Fig. (\ref{fig:2b}) shows that the atomic BEC exhibits behaviour similar to an odd cat state, as seen in optical systems. In all three solutions, the molecular BEC takes the form of droplets, being a linear superposition of kink and anti-kink, as illustrated in Eq. (\ref{psi_m_spt}). The corresponding Wigner distribution is plotted in Fig. (\ref{fig:3}) in the $x$ and $p$ quadratures, indicating squeezing.
\begin{figure}
    \centering
    \includegraphics[width=0.45\textwidth]{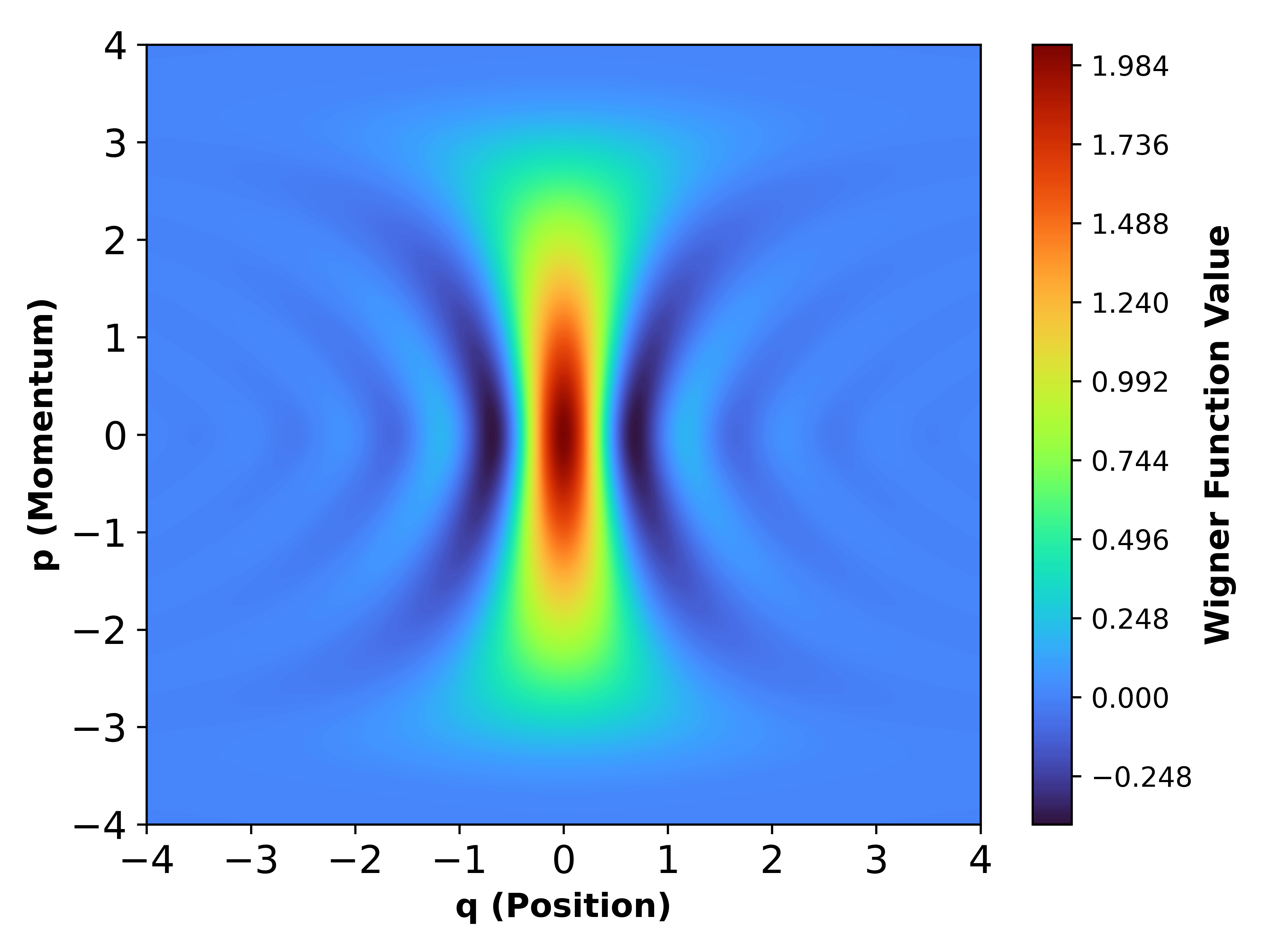}
    \caption{Plot of the Wigner function for a molecular Bose-Einstein condensate (BEC) composed of kink and anti-kink, revealing asymmetry in phase space characteristics of squeezing, with the parameters $\beta=2.1089$ and $\Delta=7.2545$.}
    \label{fig:3}
\end{figure}

\section{Conclusion and Future Scope}

In conclusion, we find exact solutions for quantum droplets and Schrödinger cat states as solutions of the mean field equations of AMBEC.
The shapes of both atomic and molecular states are governed by the chemical potential, with the droplets adopting a characteristic flat-top super-Gaussian structure due to the limiting behavior of the chemical potential. The limiting value of $\mu$ is determined by the system's nonlinearities and the atom-molecular interconversion.
We have demonstrated that odd and even atomic cat states can exist as exact eigen-solitons in the atom-molecular BEC system. These states are superpositions of bright solitons in the atomic component, arranged in a cat state configuration, while the molecular BEC forms a superposition of dark solitons, where the quadratic nature of the interconversion term plays a crucial role in the formation of cat states. It is further observed that in atomic BEC, while quadratic nonlinearity can lead to the formation of droplets, it does not result in the generation of atomic cat states. The specific nature of the interconversion term facilitates the formation of bright solitons and their superposition in the atomic condensate, where cat states manifest. Remarkably, the nonlinear Schrödinger equation with the LHY correction supports the superposition of dark solitons, while the asymmetric structure of the quadratic atom-molecular interconversion term enables the realization of both bright and dark soliton superpositions as eigen-solitons. With the recent observation of droplets in BEC systems, the detection of cat states and droplet configurations in AMBEC systems may be possible. Additionally, the interconversion and other forms of Kerr-type nonlinearities can be controlled via Raman-induced photo-association and Feshbach resonance, potentially allowing the production and manipulation of such macroscopic quantum states. Cat states hold significant value not only for foundational aspects of quantum mechanics, such as the study of wavefunction collapse, but also for practical applications in quantum computing, quantum sensing, and sub-Fourier precision measurements. {We, therefore, displayed  the Wigner functions
for the even and odd cat states, respectively in Fig. (\ref{fig:2a}) and Fig. (\ref{fig:2b}), revealing
their discernible interference behavior in phase-space, with the odd one having
vanishing intensity at the origin. The droplets being composed of kink-anti-
kink combination, display assymetric intensity distribution in coordinate and
momentum, characteristic of squeezing.

For the Gaussian form of the droplets, the harmonic oscillator potential is obtained and for the case of flat top, one obtains the box type potential. For the cat states, potentials are a double well type potential as the wavefunction is localized in two different points. 
It may be noted that in this paper we have chosen $B>0$ for all three solutions so
that one can express the three solutions as superposition of either bright solitons or as superposition of kink and anti-kink. However, strictly speaking it is enough to have $B > -1$ in all three solutions so as to have non-singular solutions. It may be 
worthwhile examining in detail the behaviour of the three solutions in case $-1 < B < 0$. We plan to examine in detail this 
case and hope report its consequences in near future. Just as the group-theoretical properties of coherent states have been extensively studied \cite{prakash1994mixed}, a similar group-theoretical analysis could be explored for atomic cat states in Bose-Einstein condensates (BECs) in future research.

\appendix 
\section{}\label{Appendix_A}
Here, we analyze the consistency conditions for all the solutions. First solution corresponds to  droplets and the second and third case corresponds to atoms in the even and odd cat states, respectively. In all of these cases the molecular BEC are in the droplet form.\\

{\bf Solution I}:

In case of droplets the field ansatz in Eq. (\ref{3})
yields the three relations
\be\label{A1}
\mu = -2\beta^2\,,
\ee
\be\label{A2}
\sqrt{2} \alpha D  = -3(1+2B)\beta^2\,,
\ee
\be\label{A3}
g_{a} A^2 + g_{am} D^2 = 4B(B+1)\beta^2\,.
\ee

On the other hand, on using the field ansatz in Eq. (\ref{4}) yields the four 
relations
\be\label{A4}
\epsilon = -3\beta^2\,,~~D = \pm A\,,
\ee
\be\label{A5}
(g_{am}+g_{m})A^2 = 2B(1+B)\beta^2\,.
\ee
This then leads to :
\be\label{A6}
(g_a - g_m) A^2 = 2B(B+1) \beta^2\,.
\ee
From the above equation, it is evident that this solution is not valid when the atom-atom interaction and molecule-molecule interaction have the same strength.\\
Straightforward calculation from the  above consistency relations leads to the
following relations:
\be\label{A7}
\epsilon = \dfrac{3}{2}\mu= -3\beta^2
\ee
\be\label{A8}
A^2 = D^2 = \left(\dfrac{\beta^2}{\dfrac{2\alpha^2}{9\beta^2} - g_a - g_{am}}\right)
\ee

Inter- and intra-species interactions between atoms and molecules are found to be related as: 
\be\label{A9}
g_m = \dfrac{g_a-g_{am}}{2}
\ee
Further, we can infer that $\mu,\epsilon$ and $D<0$ for this solution for $B>0$.

\textbf{ Solution II} 

We use the field ansatz in Eq. (\ref{4}), which yields the following relations satisfied by different relations
\be\label{A10}
(2\mu -\epsilon)D = -D\beta^2 +\frac{\alpha A^2}{\sqrt{2}}\,,
\ee
\be\label{A11}
g_{am} A^2 = (2\mu-\epsilon)B -\frac{(3+4B)}{2}\beta^2\,,
\ee
\be\label{A12}
g_m D^2 = (2\mu -\epsilon)B^2 + \frac{B}{2}\beta^2
\ee

On the other hand, on using the field ansatz in Eq. (\ref{3}) yields the following three relations
\be\label{A13}
\mu = -\frac{\beta^2}{2}\,,~~g_a A^2 + \sqrt{2} \alpha D = -(1+4B)\beta^2\,,
\ee
\be\label{A14}
g_{am} D^2 - B g_a A^2 = 4B(B+1)\beta^2\,.
\ee
From Eq. (\ref{A10}), we can find the relation between A and D:
$$A^2=-\frac{\epsilon}{\alpha}\sqrt{2}D$$
Eqs. (\ref{A11}) and (\ref{A13}) give us the following relation
$$D=\Gamma B$$
where, $$\Gamma=\frac{\sqrt{2}(\epsilon+6\mu)\alpha}{2g_{am}\epsilon-3g_a\epsilon+3\alpha^2}$$
Now Eq. (\ref{A13}) gives us 
\be \label{A15}
B=\frac{\sqrt{2}\mu\alpha}{(\alpha^2-g_a\epsilon)\Gamma-4\sqrt{2}\mu\alpha},
\ee
and Eq. (\ref{A12}) becomes
\be\label{A16}
B=\frac{\mu}{(2\mu-\epsilon)-g_m\Gamma^2},
\ee
The Eq. (\ref{A14}) can be expressed as 
\be\label{A17}
B=\frac{-8\mu\alpha}{g_{am}\alpha\Gamma^2+\sqrt{2}g_a\epsilon\Gamma+8\mu\alpha} 
\ee
Using Eq. (\ref{A15}), the Eqs. (\ref{A16}) and (\ref{A17}) can be cast into following form :
$$\sqrt{2}g_m\alpha\Gamma^2+(\alpha^2-g_a\epsilon)\Gamma-\sqrt{2}\alpha(6\mu-\epsilon)=0$$
and
$$ g_{am}\alpha\Gamma^2+\sqrt{2}(4\alpha^2-3g_a\epsilon)\Gamma-24\mu\alpha=0.$$

These two consistency conditions relate the parameters $\mu$, $\epsilon$ and $\alpha$ in a wide range of their values. Alternatively, using $\dfrac{D^2}{A^2}>0$ in Eqs. (\ref{A10}) - (\ref{A14}) we find that $\mu,  \epsilon, g_a < 0$ while $g_m, D > 0$ for this solution for $B>0$.\\

{\bf  Solution III}
 
On using the field ansatz in Eq. (\ref{4})
yields three relations
\be\label{A18}
(2\mu -\epsilon)D = -D\beta^2 +\frac{\alpha A^2}{\sqrt{2}}\,,
\ee
\be\label{A19}
g_{am} A^2 = (2\mu-\epsilon)(B+1) -\frac{(1+4B)}{2}\beta^2\,,
\ee
\be\label{A20}
g_m D^2 = (2\mu -\epsilon)(B+1)^2 - \frac{(B+1)\beta^2}{2}\,.
\ee

On the other hand, on using the field ansatz in Eq. (\ref{3}) yields the three 
relations
\be\label{A21}
\mu = -\frac{\beta^2}{2}\,,~~g_a A^2 + \sqrt{2} \alpha D = -(3+4B)\beta^2\,,
\ee
\be\label{A22}
g_{am} D^2 - (1+B) g_a A^2 = 4B(B+1)\beta^2\,.
\ee

From Eq. (\ref{A18}), we can find the relation between A and D:
$$A^2=-\frac{\epsilon}{\alpha}\sqrt{2}D$$\\
Eqs. (\ref{A19}) and (\ref{A21}) give us the following relation
$$D=\Gamma B + C$$
where,
$$\Gamma = \frac{\sqrt{2}(\epsilon-2\mu)\alpha}{2g_{am}\epsilon-g_a\epsilon+\alpha^2}$$
and $$C=\frac{\sqrt{2}\epsilon\alpha}{2g_{am}\epsilon-g_a\epsilon+\alpha^2}$$

Now Eq. (\ref{A21}) can be expressed as 
\be 
B=\frac{3\sqrt{2}\mu\alpha+(g_a\epsilon-\alpha^2)C}{(\alpha^2-g_a\epsilon)\Gamma-4\sqrt{2}\mu\alpha},
\ee

Eq. (\ref{A20}) can be expressed as

\begin{align}
     & B^2[g_m\Gamma^2 - (2\mu-\epsilon)]  + B[2g_m\Gamma C-5\mu+2\epsilon]~~ + \notag \\
     & [g_m C^2-(3\mu-\epsilon)]=0 \label{A24}
\end{align}

and Eq. (\ref{A22}) can be expressed as

\begin{align}
    & B^2[g_{am}\alpha\Gamma^2 + 8\mu\alpha + \sqrt{2}g_a\epsilon\Gamma] + 
    B[2g_{am}\alpha\Gamma C~~+ \notag \\
    &\sqrt{2}g_a\epsilon (\Gamma + C) + 8\mu\alpha] + [g_{am}\alpha C^2 + \sqrt{2}g_a\epsilon C] = 0 \label{A25}
\end{align}

Using Eq. (A23), the Eqs. (A24) and (A25) yields two additional consistency conditions as we obtained in Solution-2, involving three parameters $\mu,  \epsilon$ and $\alpha$. It is found that all the three parameters need to be non-zero for the solutions to exist. Alternatively, using $\dfrac{D^2}{A^2}>0$ in Eqs. (\ref{A18}) - (\ref{A22}) we find that  $\mu, g_m, g_{am} < 0$ while $\epsilon$ and $D$ have opposite signs for this solution for $B>0$. 

\section{}\label{Appendix_B}

In the following, we identify the self consistent potentials by casting the nonlinear Schrödinger equations as eigen value equations. Explicitly, the time dependent Eq.
 (\ref{3}) and Eq. 
 (\ref{4}) are:

\begin{eqnarray}
i\frac{\partial\psi_{a}(x,t)}{\partial t} &=& -\frac{1}{2}\frac{\partial^{2} \psi_{a}(x,t)}{\partial x^{2}} + [g_{a}|\psi_{a}(x,t)|^{2} + g_{am}\nonumber \\&& 
 |\psi_{m}(x,t)|^{2}]\psi_{a}(x,t) + \alpha \sqrt{2}\psi_{m}(x,t)\psi_{a}(x,t)^{*}, \label{B1} \nonumber\\ \\
i\frac{\partial\psi_{m}(x,t)}{\partial t} &=& -\frac{1}{4}\frac{\partial^{2} \psi_{m}(x,t)}{\partial x^{2}}+[\epsilon +g_{m}|\psi_{m}(x,t)|^{2}+ \nonumber \\ &
& g_{am}|\psi_{a}(x,t)|^{2}]\psi_{m}(x,t) +
\frac{\alpha}{\sqrt{2}}\psi_{a}(x,t)^{2} \label{B2}
\end{eqnarray}
can be written as:
\bea
\mu \phi_{a}(x) &=& -\frac{1}{2} \frac{\mathrm{d^2} \phi_{a}(x)}{\mathrm{dx^2} } + [V_a] \phi_{a}(x), \label{B3} \\
(2\mu - \epsilon) \phi_{m}(x) &=& -\frac{1}{4} \frac{\mathrm{d^2} \phi_{m}(x)}{\mathrm{dx^2} } + [V_m] \phi_{m}(x) \label{B4}
\eea

\begin{eqnarray}
& \text{Here,} \quad \psi_{a}(x,t) = \phi_{a}(x) \cdot e^{-i \mu t}, \nonumber \\
& \text{and} \quad \psi_{m}(x,t) = \phi_{m}(x) \cdot e^{-2i \mu t} \nonumber
\end{eqnarray}

and $V_a$, 
$V_m$ are the self consistent potentials  given respectively by :

\bea \label{B5}
V_{a} = g_{a} \phi_{a}(x)^{2} + g_{am} \phi_{m}(x)^{2} + \alpha \sqrt{2}  \phi_{m}(x),
\eea
\bea \label{B6}
V_{m} = g_{m}\phi_{m}(x)^{2} + g_{am}\phi_{a}(x)^{2} +
\frac{\alpha}{\sqrt{2}}\frac{\phi_{a}(x)^{2}}{\phi_{m}(x)} 
\eea

It is evident that these potentials is different for droplets and cat states as the field configurations are different. These are clearly evident in the potential figures.


\bibliographystyle{unsrt}

\end{document}